\documentclass[11pt,a4paper]{article}
\usepackage{caltn,psfig}

\begin{document}

\docnum{IACHEC Report 2015}
\title{Summary of the 10th IACHEC Meeting}
\author{Xiaobo Li$^a$, Shu Zhang$^a$, Catherine Grant$^b$, Matteo Guainazzi$^{c,d}$, Eric Miller$^b$, \\ Lorenzo Natalucci$^e$, Jukka Nevalainen$^f$, on behalf of the IACHEC \\
  ($^a$Institute of High Energy Physics, Beijing, PRC \\
   $^b$MIT Kavli Institute for Astrophysics and Space Research, Cambridge, U.S.A.\\
   $^c$European Space Astronomy Centre of ESA, Madrid, Spain \\
   $^d$Institute of Space and Astronomical Science, Sagamihara, Japan \\
   $^e$Istituto di Astrofisica e Planetologia Spaziali, INAF, Roma, Italy \\
   $^f$University of Tartu, Estonia)
}
\date{\today}                   

\maketitle

\begin{center}
{\small
{\bf Abstract} \\
We summarize the outcome of the 10th meeting of
the International Astronomical Consortium for High Energy Calibration
(IACHEC), held in Beijing (People's Republic of China) in April 2015.
Over 80 scientists
directly involved in the calibration of operational and 
future high-energy missions gathered during 3.5~days to discuss the status of
the X-ray payload inter-calibration, as well as possible ways to improve it.
A recent study on a large sample of galaxy clusters
confirmed that
the calibration of the effective area shape above 2 keV
between {\it XMM-Newton}/EPIC and
{\it Chandra}/ACIS is consistent, but showed
a significant discrepancy at lower energies.
Temperatures measured by EPIC are therefore smaller,
the
difference being largest for the highest temperatures, up to
$\simeq$20\% at $kT$=10~keV (Schellenberger et al. 2015).
The latest multi-mission study of the Crab Nebula above 10~keV
shows a $\pm$13\% agreement in the relative normalization of the
INTEGRAL, NuSTAR, RXTE, and {\it Suzaku} hard X-ray instruments.
}
\end{center}

The International Astronomical Consortium for High Energy Calibration
(IACHEC)\footnote{{\tt http://web.mit.edu/iachec/}}
is a group dedicated to supporting the cross-calibration
environment of high energy astrophysics missions with the ultimate goal
of maximizing their scientific return. Its members are drawn from
instrument teams, international and national space agencies and other
scientists with an interest in calibration in this area. Representatives
of over a dozen current and future missions regularly contribute to the
IACHEC activities. Support for the IACHEC in the form of travel costs
for the participating members is generously provided by the relevant
funding agencies.

IACHEC members cooperate within working groups to define calibration
standards and procedures. The scope of these groups is primarily a
practical one: a set of data and results (eventually published in
refereed journals) will be the outcome of a coordinated and standardized
analysis of reference sources (``high-energy standard candles'').
Past, present and future high-energy missions can use these results as a
calibration reference.
The 10th IACHEC meeting was successfully hosted by HXMT team during April
20--23 2015 at the Fragrant Hill of Beijing. It turns out to be the
IACHEC with the largest number of participants so far. 85 scientists
covering a diversity of nationalities, including China, US, Germany, Spain,
Italy, India, Japan, and Estonia, joined this meeting and shared their
experience
on calibration of operational as well as future missions.

At the end of the nominal meeting,
a mini-workshop was held at IHEP to discuss the
calibration status and plans of the Hard X-ray Modulation Telescope.
Six IACHEC participants out of China attended this mini-workshop as experts.
The event was highly successful.
It could represent a precedent to increase the efficiency and
impact of future IACHEC meetings.   

\section{Working Group reports}

\subsection{CCD}

The CCD Working Group again met in conjunction with the Backgrounds
Working Group due to substantial overlap in interested participants. As
always, the CCD Working Group provided a forum for cross-mission
discussion and comparison of CCD-specific modeling and calibration
issues, while the Backgrounds Working Group provides the same for
measuring and modeling instrument backgrounds in the spatial, spectral
and temporal dimensions.  At IACHEC 2015, we heard from {\it Chandra}/ACIS,
{\it Chandra}/ACIS/HETG, and {\it XMM-Newton}/EPIC-pn.

We started the session with presentations on ACIS calibration and
background. Nick Durham summarized work using the ACIS external
calibration source (radio-isotope Fe-55 plus Al and Ti fluorescence
targets).  He is evaluating a variety of calibrated quantities as a
function of time and focal plane temperature, such as line centroid and
width, quantum efficiency uniformity across the CCDs, and the spatial
distribution of the contamination on the ACIS filters.  Terry Gaetz
spoke on efforts to improve the accuracy of the low-energy gain on the
ACIS BI CCDs using Low-Energy Transmission Gratings spectra, and also on
a preliminary investigation into ACIS background flares and their
frequency.  Improvements for the ACIS-S1 low-energy gain based on this
work will be released shortly in CALDB 4.6.9. He also found that while
the fraction of time during background flares is higher during solar
maximum than minimum, the current solar maximum has a much lower flaring
fraction than the previous one.

Matteo Guainazzi and Norbert Schultz both gave presentations on a
multi-observatory calibration effort to better understand the EPIC-pn
response to bright sources at low energies and more generally, CCD
redistribution in specialized timing modes.  {\it Chandra}/HETG,
{\it XMM-Newton}, and
{\it Swift} all simultaneously observed the bright obscured X-ray binary Cyg
X-3.  The {\it Chandra} HETG line energies can then be fed back into the CCD
spectra to better calibrate the energy scale, while comparison of low
energy residuals can feed into improved low energy redistribution
models.

Finally, Michael Smith discussed the EPIC-pn energy scale and the
quiescent background. After adjusting event energies with the long-term
CTI correction calibrated using the on-board calibration source, there
is a residual secular trend that is strongly correlated with the
quiescent background. The residual secular trend is not due to CTI, but
is a background-dependent gain effect. SAS implementation and scientific
validation of an additional background-dependent gain correction is in
progress.

\subsection{Contamination}

The contamination Working Group held its second face-to-face meeting at the 
2015 IACHEC workshop, gathering 6 of 20 WG members and
several additional non-members for a single session.
The group discussed the affects of molecular contamination 
(Marshall et al. 2004, Koyama et al. 2007, O'Dell et al., 2013)
on soft X-ray ($\leq$ 1 keV) instruments in light of the three broad topics
introduced at the inaugural meeting in 2014: \\
(1) comparison of contamination among instruments and missions; \\
(2) mitigation for current instruments; and \\
(3) mitigation for future instruments. 

Representatives from operating missions presented status reports on the
contamination in their soft X-ray CCD instruments.  Herman Marshall
presented new observations of the blazar Mrk 421 with the {\it Chandra} LETG and
ACIS-S array.  This ``big dither'' observation was similar to one performed
in 2014, where the pointing direction was dithered by $\pm$64~arcsec in a
direction orthogonal to the grating dispersion direction.  This dither is
much larger than the $\pm$8~arcsec Lissajous dither pattern in standard
observing, and allows fine mapping of the contaminant spatial distribution
along this direction.  The results indicate that the contamination is
continuing to build at an increasing rate on ACIS, and appears symmetric
top-to-bottom in the ACIS-S array, with more contaminant at the edges of
the filter.  The oxygen to carbon ratio varies from the center to edge,
with possibly two contaminant components responsible.  Doug Swartz
presented an updated model of contamination migration within {\it Chandra},
building on previous work constraining the volatility of the contaminants.
These results suggest that most of the recently accumulated contamination
comes from a second, higher volatility source that has become increasingly
active,  possibly due to rising temperatures within the {\it Chandra}
Observatory.  If the source rate continues to track these temperature
trends, then the contamination layer might dramatically increase in the
near future.  Eric Miller showed updates of the {\it Suzaku} XIS contamination
monitoring, using targets such as SNR 1E~0102.2$-$7219, the isolated
neutron star RX~J1856.5$-$3754, and the BLLac PKS~2155$-$304.  The XIS
show no evidence for an accelerated source of contaminant as with {\it Chandra},
and in fact have continued to show signs of decreasing contamination
optical depth.  The total effective area at 0.65 keV (the O VIII Ly$\alpha$
line) has been increasing at about 10\% per year since 2011.

For upcoming missions, Kallol Mukherjee briefly discussed the possible
sources of contamination in the Soft X-ray Telescope (SXT) CCD instrument
aboard ASTROSAT, along with plans for initial operations to mitigate early
contaminant build-up.  There was significant discussion among the working
group members about how long to wait before opening the instrument door, what
initial targets to point at for a zero-contamination baseline, and what
might be expected given the CCD operating temperature and other operational
parameters.

Efforts will continue before the next IACHEC meeting to work toward a
legacy white paper detailing shared lessons learned about contamination on
soft X-ray instruments.

\subsection{Galaxy Clusters}

\subsubsection{Multi-Mission Study}

We continued the discussion, started in the previous meeting, about the 
”Multi Mission Study” project, which currently consists of comparing X-ray 
spectroscopic results of four clusters obtained with on-going and past X-ray 
missions/instruments {\it XMM-Newton}/EPIC, {\it Chandra}/ACIS, {\it Swift}/XRT, {\it Suzaku}/XIS and 
ROSAT/PSPC. 

While we do have preliminary results indicating significant patterns in the 
behavior of the cross-calibration, we need help with interpreting them. Our 
aim is to utilize the collective IACHEC experience on the involved instruments 
and their calibration. We challenge anyone interested to respond to our related 
``Request to the IACHEC community'' found in the project
page\footnote{{\tt https://wikis.mit.edu/confluence/display/iachec/MMS}}
We aim for a refereed publication on the topic before the next IACHEC meeting.

We discussed about the possibility of extending the ``Multi Mission Study'' 
cluster data base to include NuSTAR mission and the near-future Astro-H, 
Astrosat and eROSITA missions. This data base could be very valuable for the 
0.5-10 keV band effective area calibration of the future missions. We are 
currently negotiating with the calibration teams of the above missions about 
including our cluster sample as calibration targets. If this turns out to be 
infeasible, we will proceed via Guest Observer programs.  

\subsubsection{Stack residuals ratio}

Our basic tool for evaluating the effective area cross-calibration 
uncertainties is the stack residuals ratio method (e.g. Kettula et al., 2013; 
Schellenberger et al., 2015). The method is applicable to other astronomical 
targets as well. We started discussions about applying it to the blazar data 
obtained via simultaneous {\it XMM-Newton} and {\it Chandra} observations (Smith \&
Marshall, in preparation). If we find similar {\it XMM-Newton}/{\it Chandra} 
cross-calibration uncertainties using galaxy clusters or blazars, as expected, 
we will yield additional confidence on our methods. This in turn would result
in a larger statistical sample for the cross-calibration work and thus better 
precision for the calibration. 

\subsubsection{HIFLUGCS follow-up}

We discussed a possible follow-up cross-calibration work using the 
{\it XMM-Newton}/{\it Chandra} HIFLUGCS cluster sample (Schellenberger et al., 2015, see 
also the
catalog\footnote{{\tt http://vizier.u-strasbg.fr/viz-bin/VizieR?-meta.foot\&-source=J/A\%2bA/575/A30}}). The larger sample of hot clusters and the
additional exposure time, compared to that in Nevalainen et al., (2010) could 
be used to improve the Fe XXV/XXVI emission line ratio diagnostics. The line 
ratio measurement yields the estimate of the ionization temperature which, due 
to the very narrow energy band used, is independent on the accuracy of the 
effective area shape calibration. Comparison with the continuum-based 
bremsstrahlung temperature (affected by the effective area calibration 
accuracy) yields an additional tool for estimating the calibration accuracy. 
For cool clusters there is also a possibility of using the emission lines of 
sulfur (S~XV and S~XVI) and silicon (Si~XIII and Si~XIV) for the temperature 
measurements. 

\subsubsection{Cluster X-ray and gravitational lensing masses as calibrators}

Inspired by an IACHEC-related paper (Israel et al., 2015), we discussed the 
comparison of cluster masses obtained with X-rays with those obtained with 
gravitational lensing. Unfortunately the situation has not improved from what 
we reported last year; the possible existence of non-thermal pressure in 
clusters would induce a hydrostatic bias with similar effect on the X-ray 
masses as caused by the {\it XMM-Newton}/{\it Chandra} effective area cross-calibration 
uncertainties. A possible near-future solution could be achieved with Astro-H 
measurements of the non-thermal component via the broadening of the Fe~XXV 
emission lines in clusters.

\subsection{Heritage}

This Working Group aims at preserving the IACHEC corpus of knowledge,
know-how and best practices for the benefit of future missions and the
community at large. Its main goals are:

\begin{itemize}

\item provide a platform for the discussion of experiences coming from
  operational missions
\item  facilitate the usage of good practices for the management of pre- and
  post-flight calibration data and procedures, and the maintenance and
  propagation of systematic uncertainties (the latter task in strict
  collaboration with the "Calibration uncertainties" IACHEC Working Group)
\item document the best practices in analyzing high-energy astronomical data
  as a reference for the whole scientific community
\item ensure the usage of homogeneous data analysis procedures across the IACHEC
  calibration and cross-calibration activities
\item consolidate and disseminate the experience of operational missions on the
  optimal calibration sources for each specific calibration goal

\end{itemize}

The activity for the following term will concentrate on:

\begin{itemize}

\item publishing a paper on the in-flight calibration plans of currently
  operational mission on the Journal of Astronomical Telescope, Instruments and
  Systems (Guainazzi et al., in preparation)

\item producing a summary of ``best-practices'' as far as
  a) photoelectric absorption models and associated cross- sections;
  b) elemental abundance tables;
  c) optically thin equilibrium emission plasma codes, and benchmark the
  effects (if any) of different prescriptions [a)+b)] on the calibration
  results

\item populating the repository of calibration documents on the WG
  Wiki\footnote{{\tt https://wikis.mit.edu/confluence/display/iachec/IACHEC+Heritage+Working+Group}}

\item build a “IACHEC knowledge database”

\end{itemize}

\subsection{Non-thermal SNRs}

A meeting of the Non-thermal SNR group was held, attended by  
calibration scientists from different missions with a significant  
share of Chinese representatives. The attendees list is the following:  
Juan-Ha Wang, Haihui Zhao, Xin Zhou, Dipankar Bhattacharya, Yoshitomo  
Maeda, Michael Smith, Fangjun Lu, Paul Plucinsky, Xi Long, Lorenzo  
Natalucci, Mingyu Ge. In addition, Kristin Madsen participated  
remotely (Skype).

The group addressed first the cross-calibration on the hard X-ray  
band, using the Crab Nebula as a reference source. This effort  
includes the analysis of data from {\it Suzaku} (XIS,HXD), RXTE (PCA)
{\it XMM-Newton}  
(EPIC-pn), INTEGRAL (IBIS,SPI) and NuSTAR. Updates since year 2014  
include new assessments of calibrations for NuSTAR (Madsen et al.  
2015) and for IBIS/ISGRI (Savchenko et al. 2014). A simultaneous  
{\it XMM-Newton}/NuSTAR/INTEGRAL observation took place in November 2014.
For {\it XMM-Newton},  
the data were taken from EPIC-pn in burst mode and a possible update  
of the burst mode calibration will follow.

Progress on the analysis was achieved by the addition of new data sets  
from most recent periods, yielding a total of 11 nearly simultaneous  
epochs up to November 2014. The new observations also include data from  
NuSTAR. The multi-instrument fits from single epochs show evidence of  
a break at high energies (near 100 keV) by SPI and HXD/GSO. The HXD  
normalizations are generally found to be higher than INTEGRAL ($\simeq$13\%  
and 8\% for the PIN and GSO, respectively), while the PCA lies in  
between ($\simeq$6\%). The NuSTAR normalization is significantly lower than  
the INTEGRAL one ($\simeq$12\%).

The results obtained from the Crab are broadly consistent with the  
ones reported by Tsujimoto et al. (2011) from observations of  
G21.5-0.9. However, the relative spread in the measured flux values is  
significantly wider in the case of G21.5-0.9. For this source, a break  
at 9 keV was reported by NuSTAR (Nynka et al. 2014), which is the  
possible explanation for the above discrepancy. It was noted that in  
the Tsujimoto et al. paper a clear difference emerges between values  
of the spectral slope measured independently in the soft and hard band.
A natural explanation of this effect would be the presence of the 9  
keV spectral break, and not an instrumental bias. A test of the new  
model against {\it Chandra}, NuSTAR and INTEGRAL is planned.

Finally, Mingyu Ge presented new results for Crab timing mostly  
obtained from analysis of RXTE data. The analysis shows clear evidence  
for a phase lag between the X-ray and radio pulses, increasing with  
time. M. Smith presented results obtained by
{\it XMM-Newton}. A discussion  
followed on whether it could be possible to use other instruments data  
(e.g. INTEGRAL) to confirm these results.

\section{Cross-calibration status}

The latest IACHEC
cross-calibration study was published by Schellenberger et al.
(2015). These authors
discuss an update of the cross-calibration status between the
{\it Chandra}/ACIS and the {\it XMM-Newton}/EPIC, based on
a large sample of galaxy clusters. This study
confirms previous results
in, {\it e.g.}, Nevalainen et al. (2010).
The calibration of the effective area shape above 2 keV
between EPIC and ACIS is consistent, but there is a significant problem at
lower energies resulting in significantly smaller EPIC temperatures.
The temperature
difference is bigger for the highest temperatures, amounting to
$\simeq$20\% at $kT$=10~keV.
Readers are refereed to Sect.~2
of Burrows et al. (2014) for a summary of the cross-calibration
status among operational missions based on samples of
point-like sources (cf. their Fig.~4).

\section*{References\footnote{see {\tt http://web.mit.edu/iachec/papers/index.html} for a complete list of IACHEC papers}}

\noindent
Burrows et al., 2014, IACHEC Report
2014\footnote{available at {\tt http://arxiv.org/pdf/1412.6233}}\\
\noindent
Kettula et al., 2013, A\&A, 552, 47 \\
\noindent
Koyama et al., 2007, PASJ, 59, 23 \\
\noindent
Israel et al., 2015, MNRAS, 448, 814 \\
\noindent
Madsen et al. 2015, ApJ Suppl., 220, 8 \\
\noindent
Marshall et al., 2004, in Society of Photo-Optical
 Instrumentation Engineers (SPIE) Conference Series, Vol. 5165, X-Ray and
 Gamma-Ray Instrumentation for Astronomy XIII, ed. K.~A.~Flanagan \&
 O.~H.~W.~Siegmund, 497--508 \\
\noindent
Nevalainen et al., 2010, A\&A, 523, 22 \\
\noindent
Nynka et al. 2014, ApJ 789, 72 \\
\noindent
O’Dell et al., 2013, in Society of Photo-Optical Instrumentation Engineers (SPIE) Conference Series, Vol. 8859, UV, X-Ray, and Gamma-Ray Space Instrumentation for Astronomy XVIII, ed. O. H. W. Siegmund, 88590F \\
\noindent
Savchenko et al. 2014, Proceedings of Science, PoS(Integral2014)083 \\
\noindent
Schellenberger et al., 2015, A\&A. 575, 30 \\
\noindent
Tsujimoto et al. 2011, A\&A 525, 25 \\

\end{document}